\documentclass[a4paper]{jpconf}
\usepackage{graphicx}
\usepackage{xcolor}
\usepackage{upgreek}
\usepackage{mathtools}
\DeclareUnicodeCharacter{2212}{-}
\begin{document}
\title{Motivation and progress on superradiant active optical atomic clocks}

\author{Martina Matusko and Marion Delehaye}

\address{Université de Franche-Comté, SUPMICROTECH, CNRS, Institut FEMTO-ST, F-25000 Besançon, France}

\ead{marion.delehaye@femto-st.fr}

\begin{abstract}

Current state-of-the-art frequency standards are passive optical atomic clocks where the frequency of an optical resonator is stabilized to a narrow atomic transition. Passive clocks have achieved unprecedented stabilities of $6.6 \times 10^{−19}$ over one hour of averaging time \cite{Oelker_2019_stab_10t17}. However, they face intrinsic limitations, particularly due to thermal and mechanical fluctuations of the local oscillator. To surpass the limitations of the passive clocks and go beyond the state-of-the-art, the idea of building active optical atomic clocks emerges. These clocks would be optical counterparts of hydrogen masers, with the emitted frequency defined by the atomic transition and therefore inherently stable against cavity instabilities. This paper discusses the latest developments and future prospects in the field of active optical atomic clocks. 
\end{abstract}

\section{Introduction}
% put here the *current* applications . you can describe a bit more some experiments (geodesy, fundamental constants?)
Optical atomic clocks, with their remarkable stability~\cite{Oelker_2019_stab_10t17,McGrew_2018_geodesy_below_cm} and accuracy at the $10^{-18}$ level \cite{Nicholson_2015_uncert_st10t18, McGrew_2018_geodesy_below_cm, Brewer_2019_uncert_below_10t18}, have initiated a new era in precision measurement. This exceptional level of precision paves the way for a wide array of diverse applications, ranging from probing variations in fundamental constants \cite{ Dzuba_2000_fund_const, Godun_2014_time_var_fund_const, Huntemann2014, Safronova_2019_fund_const}  to the redefinition of the second \cite{Gurov_2013_OLC_redef_SI, Riehle_2015_redef_SI, Bregolin_2017_OLC_redef_SI, Lodewyck_2019_redef_SI, McGrew_2019_redef_SI} 
or advancing the field of the relativistic geodesy~\cite{chou_2010_opt_clocks_and_relativity, Bondarescu_2015_grav_redshift, Flury_2016_rel_geodesy, Lion_2017_geopot_model_clock_comp, Mehlstäubler_2018_clocks_4_geodesy, Grotti_2018_geodesy_transp_opt_clock, Tanaka_2021_grav_redshift}, where optical atomic clocks can serve as extremely accurate altimeters.
 %advancing the field of gravitational geodesy \cite{Grotti_2018_geodesy_transp_opt_clock, Flury_2016_rel_geodesy, Mehlstäubler_2018_clocks_4_geodesy, Lion_2017_geopot_model_clock_comp}. 
 The capabilities of these advanced timekeeping devices represent a significant leap forward, offering novel opportunities and insights across various scientific domains. 
 %Reaching the stability of $4.8 \times 10^{-17}\sqrt{{\tau}} \mathrm{s}$ and uncertainty of $2.0 \times 10^{-18}$ \cite{Oelker_2019_stab_10t17, Bothwell_2019_JILA_OLC_uncert}, optical atomic clocks have proven immensely valuable across a variety of applications. 
 Yet, surpassing the current state-of-the-art clocks in stability and accuracy would open even more advanced possibilities, especially in seismic mitigation studies, deep-space navigation applications and fundamental sciences.

The development of optical atomic clocks with fractional frequency stabilities below $10^{−18}$ achieved within an hour is for instance crucial for measuring height differences below 1 cm accuracy. This precision surpasses that of GNSS monitoring, which offers a best-case height uncertainty of typically a few cm due to atmospheric delays~\cite{Choy2017_GNSS}, with longer integration time. The ongoing improvements in optical clocks are expected to significantly enhance seismic and volcanic hazard mitigation studies \cite{Tanaka_2021_grav_redshift}. 
Essential for realizing the full potential of this technology are stable and accurate on-field systems or strong fiber links between optical atomic clocks.
%, enabling quicker and more precise detection of Earth's crust movements and transforming our approach to tectonic activity.

%%This technological progression will enable more precise and rapid detection of the earth's crust movements, offering a transformative approach in understanding and responding to tectonic activities. 

%This enhanced precision is important not only for improving terrestrial measurements but also for enabling advanced applications in space-based navigation \cite{Ely_2018_deep-space-nav}, where the Deep Space Atomic Clock (DSAC) offers \textcolor{purple}{can offer} unprecedented accuracy in one-way radiometric tracking, facilitating autonomous navigation and improving the efficiency of spacecraft operations in deep space missions.

The improved precision of atomic clocks not only enhances terrestrial measurements but also opens up advanced possibilities in space-based navigation. The Deep Space Atomic Clock (DSAC), a mercury-based ion clock, with its demonstrated stability of $2\times10^{-15}$ in one day of averaging time was operated between 2019 and 2021. Even though it is a microwave clock, it already offered unprecedented accuracy in one-way radiometric tracking, improving the efficiency of spacecraft operations in deep space missions \cite{Ely_2018_deep-space-nav}, showing potential future improvements of space studies that would be enabled by more performant space clocks~\cite{Schkolnik_2023}.  

Current bounds on fundamental constants have been established by today's best optical clocks and do not show yet any clear variation. 
%Current advancements in optical atomic clocks have made it possible to establish bounds on the variations of fundamental constants. 
A network comprising various types of clocks, including highly-charged ion clocks, molecular, and nuclear clocks, offers a promising strategy for measuring variations in the fine structure constant ($\alpha$) and the electron-to-proton mass ratio ($\mu$) with unparalleled sensitivity \cite{Barontini_2022_fund_const_clock_network}. The association of these diverse clocks, each sensitive to variations in different ways, is expected to significantly boost the ability to detect subtle changes in these fundamental constants.
% Barontini_2022_fund_const_clock_network (this requires a network + new clocks = highly charged clocks + molecular + nuclear clocks, because they have different dependencies to constants variations) + Safronova_2019_fund_const

Furthermore, atomic clocks are sensitive to ultralight scalar bosonic dark matter, which could cause oscillations in the fundamental constants, detectable through variations in clock transition frequencies \cite{Safronova_2019_fund_const}. This sensitivity extends across a broad range of dark matter masses and interaction strengths. Networks of clocks could potentially detect transient changes in fundamental constants induced by dark matter objects with large spatial extents, such as stable topological defects \cite{Derevianko_2016_dark_matter, Wcislo_2016_dark_matter_opt_clocks, Roberts_2020_dark_matter_fund_const}.
%%%MM check these three papers to make sure they are cited well

%% An array of optical atomic clocks, with a fractional timing precision of 10−18, presents a novel method for gravitational wave detection. 
%%An array of optical atomic clocks presents a novel method for gravitational wave detection \cite{loeb_2015_grav_waves}. This approach offers an alternative to the distance variation measurements used in current interferometric techniques like Advanced-LIGO and eLISA \textcolor{purple}{cite ?}. The method is distinct in that it focuses on measuring variations in timing rather than distances, aiming to detect the differential time dilation experienced by clocks located at different phases of a passing gravitational wave. 
Additionally, an array of optical atomic clocks could serve as an innovative technique for gravitational wave detection \cite{Kolkowitz_2016_grav_wave_opt_latt_clock, loeb_2015_grav_waves}. This method, contrasting with the distance variation measurements of interferometric techniques used in projects like Advanced-LIGO and eLisa, concentrates on tracking timing variations to detect the differential time dilation experienced by clocks located at different phases of a passing gravitational wave. This approach could lead to the detection of gravitational waves at various frequencies, broadening the range of perceptible astrophysical phenomena.
%not only aids in detecting gravitational waves but also offers the potential to investigate various frequencies, thus broadening the scope for studying diverse astrophysical phenomena.  

 %\textcolor{blue}{MOVE THIS TO NEXT PARAGRAPH and detecting subtle phenomena like dark matter \cite{Wcislo_2016_dark_matter_opt_clocks, Derevianko_2016_dark_matter, Wcislo_2018_dark_matter_opt_clock_network, Roberts_2020_dark_matter_fund_const}}
%\textcolor{blue}{and gravitational waves \cite{Armstrong_2006_grav_wave, loeb_2015_grav_waves},}

\begin{figure}
    \begin{center}
    \includegraphics[width=10cm]{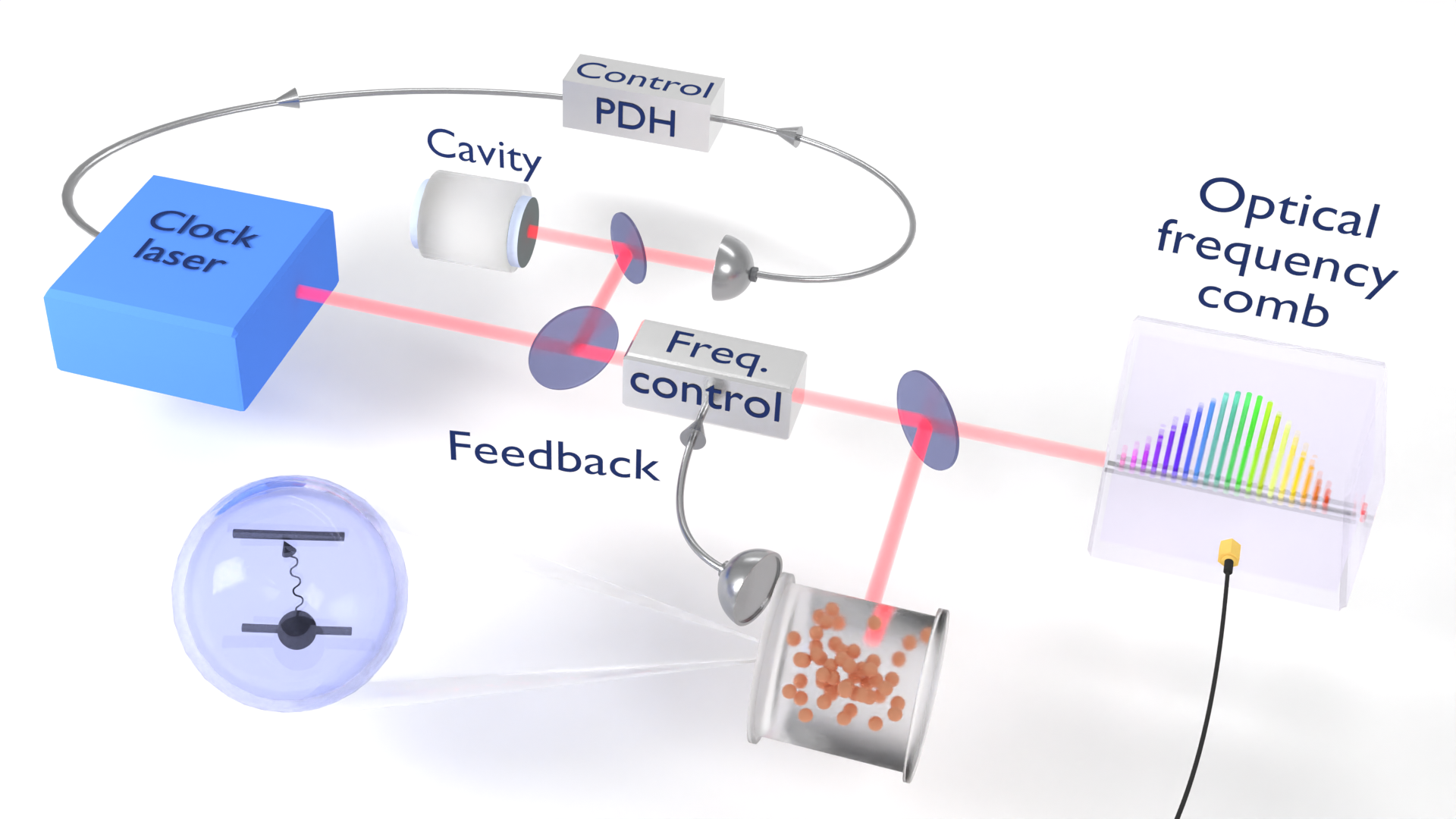}
    \caption{Scheme of a passive optical atomic clock. A clock laser, prestabilized on an ultra-stable Fabry-Perot cavity, is sent to an atomic reference, either a single ion or neutral atoms trapped in an optical lattice. }
    \label{fig_passive}
    \end{center}
\end{figure}

Current best optical clocks are based on a passive clock scheme, in which the frequency of an external local oscillator is stabilized to a narrow atomic transition. The schematic of the passive clock is depicted in Fig.~\ref{fig_passive}. 
%\textcolor{orange}{Here you wanted to mention QPN and Schawlow Townes formula, but I have no idea how to nicely incorporate it in the text since we're speaking mainly about limitations due to the LO.} 
For classical atomic ensembles, the stability of these clocks is limited by the quantum projection noise of the measurement, given by the formula: 
\begin{equation}
    \sigma_y(\tau) = \frac{1}{SNR}\frac{\Delta \nu}{\nu_0}\sqrt{\frac{T}{N\tau}},
\end{equation} 
where $SNR$ is the signal-to-noise ratio, $\Delta \nu$ is the linewidth of the transition, $\nu_0$ the transition frequency,  $T$ the interrogation time, $N$ the number of atoms and $\tau$ the integration time. 
However, this ultimate stability is not fully realized even in cutting-edge optical atomic clocks, primarily due to residual instabilities from the local oscillator. 
%%thermal noise from the reference cavity mirrors. This noise affects laser stabilization, particularly during the intervals between successive interrogation cycles.
%%Significant efforts are directed towards enhancing reference cavities to boost the overall stability of passive clocks. Alternatively, building an active optical atomic clock presents a promising approach. In that scheme, light emitted directly from a narrow atomic transition is harnessed as a precise frequency reference. Unlike passive clocks, active clocks are immune to cavity noise. Active clocks could potentially serve as an external reference for their passive counterparts. 
Here, we will provide an overview of passive optical atomic clocks and their limitations. Following this, we will focus on the new concept of active optical atomic clocks where light emitted directly on a narrow atomic transition is harnessed as a stable frequency reference. %will be discussed, outlining their current achievements, and discussing potential limitations as well as future prospects in this field.

\section{Passive optical atomic clocks}
 Current passive optical atomic clocks are realized by the frequency stabilization of an external optical local oscillator to an atomic transition. 
%Achieving an extremely high level of accuracy in frequency reference requires identifying a transition that is minimally influenced by external disturbances.
%Additionally, the choice of a narrow linewidth is crucial, as it leads to the lower fundamental limit and consequently decreases uncertainty. 
The operational cycle of a passive optical atomic clock \cite{Ludlow_2015_opt_at_clocks} involves an initial process of laser cooling and preparation of the atomic state, followed by the interrogation stage, and finally destructive state read-out atom imaging, and signal processing. 
In the state preparation stage, atoms are typically loaded into a single-ion trap \cite{Zhang_2017_cooling_ion_clock, Delehaye_2018_ion_clock, King_2022_highly_charged_ion_clock} or an optical dipole trap that follows a magneto-optical trap \cite{Takamoto_2005, Campbell_2011_ultracold_atoms_time_stand, Poli_2013, Ludlow_2015_opt_at_clocks}. 
To ensure frequency stability of the laser between two interrogations of the clock transition, the interrogating laser is frequency-stabilized onto a resonant mode of an ultra-stable Fabry-Perot cavity. 
This narrows the linewidth of the probing laser and enhances the mid-term stability by maintaining the laser close to resonance between two interrogations. 

The clock frequency can be influenced by various effects stemming from the atomic environment or operational conditions. Such influences include laser-induced electric fields \cite{Bloom_2014_nature}, blackbody radiation \cite{Falke_2014_BBR, ushijima2014cryogenic}, magnetic fields \cite{Lu_2020_Zeeman_shift_evaluation_lattice, King_2022_highly_charged_ion_clock, tang_2023_BBR_ZSS_suppression}, Doppler effect \cite{Bergquist_1987_recoil_cancellation, Diedrich_1989_sideband_cooling_Doppler_cancellation, Berkeland_1998_micromot_min_ion, Degenhardt_2005_chirped_pulses_cold_atoms, Keller_2016_ion_freq_shifts} and other systematic shifts \cite{Takamoto_2005, Baillard_2007_opt_latt_bosons, Ludlow_2008_Sr_lattice_Ca_clock_evaluation, Lemke_spin0p5_lattice_clock, Falke_2011_87Sr_opt_freq_stand}.
A comprehensive discussion of these systematic effects can be found in \cite{Ludlow_2015_opt_at_clocks, Poli_2013}. Significant efforts have been dedicated to minimizing these shifts, resulting in the enhanced accuracy of atomic clocks that is now at the low $10^{-18}$ level~\cite{Nicholson_2015_uncert_st10t18,Huntemann2016,ushijima2014cryogenic,HuangCaCryo2022,Bothwell_2019_JILA_OLC_uncert}. 
%. Such endeavors have been successful, achieving an exceptional accuracy \textcolor{blue}{of} $2 \times 10^{-18}$ level in atomic clocks \cite{Bothwell_2019_JILA_OLC_uncert}. 
%Effectively managing the system perturbations enhances both accuracy and stability.

Despite the unparalleled precision measurements made possible by passive optical atomic clocks, intrinsic limitations persist, both in terms of stability and accuracy. Even better accuracies could be obtained in principle by using highly charged ions or a nuclear transition in thorium, but limitations of the fractional frequency stabilities will be detailed in the next paragraph.
 
\subsection*{Fractional frequency stability limitations of a passive optical atomic clock}

In passive clock schemes, the laser is stabilized to the atomic transition, with the laser frequency adjusted to the atomic resonance transition through a feedback loop. 
The destructive nature of state read-out imaging necessitates thorough preparation of the subsequent atomic ensemble at each new measurement cycle.
 Given the inherently complex nature of atom preparation and the necessity for repeated cycles, only a fraction (typically 20--50\%) of the clock cycle allows for the actual probing of atoms.

Intermittent interrogation of the atoms during each measurement cycle leads to the Dick effect \cite{Dick_1990, Santarelli_1998, Quessada_2003} - an aliasing of the local oscillator high-frequency noise into the low-frequency noise within the atomic resonator bandwidth, limiting the stability of the atomic clock. 
Efforts to mitigate the Dick effect involve increasing the duty cycle, facilitating non-destructive measurements \cite{robinson_2022_nondistr_img}, and employing more stable lasers. %The improvement of the laser's stability is therefore often achieved by enhancing the quality of optical cavities.

One approach for minimizing the instability imposed by the Dick effect involves increasing the measurement duty cycle. A zero-dead-time optical clock based on the interleaved interrogation of two cold-atom ensembles demonstrated minimal Dick noise, achieving an exceptional fractional frequency instability assessed at $6 \times 10^{-17}/{\sqrt{{\tau}}}$ where $\tau$ is the integration time in seconds~\cite{Schioppo_2017_2_ensembles_clock_reduced_dick_eff}. 
Similarly, the Dick effect can be reduced when two distinct clocks operating on the same transition are compared \cite{Takano_2016_2_clocks_reduced_Dick_eff}.

%The improvement of the laser's stability aiming for the reduction of the Dick effect can be achieved by enhancing the quality of optical cavities. 
Enhancing the quality of optical cavities can lead to improved laser frequency noise and thus reduce the Dick effect. 
Presently, the most advanced optical resonators exhibit fractional frequency instabilities as low as $4 \times 10^{-17}$ \cite{Matei_2017_cavity_state_of_the_art}. However, these resonators face limitations imposed by thermal Brownian noise originating from their components, particularly mirror coatings \cite{Numata_2004_thermal_limit_cavities}. 
Their enhanced performance can be credited to several techniques developed to reduce thermal effects in optical cavities. 
One strategy for mitigating the thermal effects in the optical cavities involves the use of a long resonator operating at room temperature, simplifying the system and its operation. 
The achieved stability closely approached the thermal noise limit, highlighting the potential of long resonators at ambient temperatures for high precision in frequency stability \cite{Haefner_2015_long_cavity}. 
Moreover, significant attention is directed towards optical coating technology to tackle Brownian noise in cavity mirrors. A notable accomplishment is a tenfold reduction in Brownian noise achieved through high-reflectivity coatings formed by directly bonding monocrystalline multilayers \cite{Cole_2013_tenfold_reduction_brownian_noise}. 
Another noteworthy strategy involves operating optical cavities at cryogenic temperatures to mitigate thermal noise-induced fluctuations in the cavity's optical length~\cite{Kessler_2011_sub40mHz_laser_sicav,Zhang_2017_cryo_cav, Robinson_2019_cryo_cav}.

%Finally, the use of more stable reference cavities plays a crucial role in diminishing the Dick effect by facilitating longer probe times. Reducing the amount of time when atoms are not probed substantially enhances the system's overall stability.

%Another strategy for mitigating the instability imposed by the Dick effect involves increasing the measurement duty cycle. A zero-dead-time optical clock based on the interleaved interrogation of two cold-atom ensembles demonstrated minimal Dick noise, achieving an exceptional fractional frequency instability assessed at $6 \times 10^{-17}/{\sqrt{{\tau}}s}$ for the specified averaging time \cite{Schioppo_2017_2_ensembles_clock_reduced_dick_eff}. Similarly, the Dick effect can be reduced when two distinct clocks operating on the same transition are compared \cite{Takano_2016_2_clocks_reduced_Dick_eff}.
 
%Additionally, another study showed that by analyzing the frequency difference between two uncorrelated regions of the same atomic ensemble, the fractional frequency instability of a passive optical atomic clock could be as low as $4.4 \times 10^{-18} / \sqrt{\tau[s]}$ \cite{Bothwell_passive_stab_4times_10_18_two_regions_one_ensemble}.

The challenges associated with the Dick effect are therefore reaching the technological limits of feasible improvements in cavity design. Hence, there is a need for new approaches. One proposed method to enhance the properties of the optical local oscillator involves employing ``Spectral Hole Burning" spectroscopy \cite{Cook_2015_SHB, Thorpe_2011_SHB}. In this method, narrow spectral holes are imprinted into a cryogenically cooled non-linear crystal. These features are then employed for laser-frequency stabilization, where the laser frequency is actively tuned to maximize transmission through the crystal. 
%A promising fractional frequency stability of $10^{-15}{\tau}^{-1/2}$  has been experimentally demonstrated \cite{Cook_2015_SHB, Thorpe_2011_SHB}, but the necessary cryogenic setup is hardly compatible with applications outside the laboratory. Another proposition that involves a passive-like clock scheme is Ramsey-Bordé matter-wave interferometry emerges as a strong candidate for enhancing laser stabilization, demonstrating fractional frequency instability below $2\times10^{-16}$ in the short term \cite{Olson_2019_RBI}. 
The proposal of employing Ramsey-Bordé matter-wave interferometry within a passive-like clock scheme has emerged as another strong candidate for advancing laser stabilization~\cite{Olson_2019_RBI}. This approach has demonstrated fractional frequency instability below $2\times10^{-16}$ for short integration times.  
Utilizing thermal atomic beams, this method significantly outperforms the stability of other thermal atomic systems and requires minimal thermal shielding and vibration isolation, potentially surpassing the Fabry-Perot cavity in terms of insensitivity to thermal and mechanical disturbances. Demonstrated instability results, combined with advantages in low vibration sensitivity and compact design, position Ramsey-Borde optical-frequency stabilization as a highly promising approach for diverse applications. However, due to the pulsed operational nature, the standard limitations due to the Dick effect persist. 

Proofs of principle for vapor-cell based optical frequency references have also been demonstrated, using either a two-photon transition in ${}^{87}\mathrm{Rb}$~\cite{Newman2019_integrated_opt,Newman2021_two_phot} or dual-frequency sub-Doppler spectroscopy in ${}^{133}\mathrm{Cs}$~\cite{Gusching2023}. These experiments continuously interrogate the atoms and do not require an optical Fabry-Perot cavity, they are therefore not sensitive to the Dick effect. However the achieved stability has been so far limited to the $10^{-13}$ range at one second integration time. 

In the rest of this article, we will focus on a proposal for an active optical atomic clock that appears promising in overcoming these Dick effect-related limitations.

\section{Active optical atomic clocks} 
\subsection{Generalities}

\begin{figure}
    \begin{center}
    \includegraphics[width=8cm]{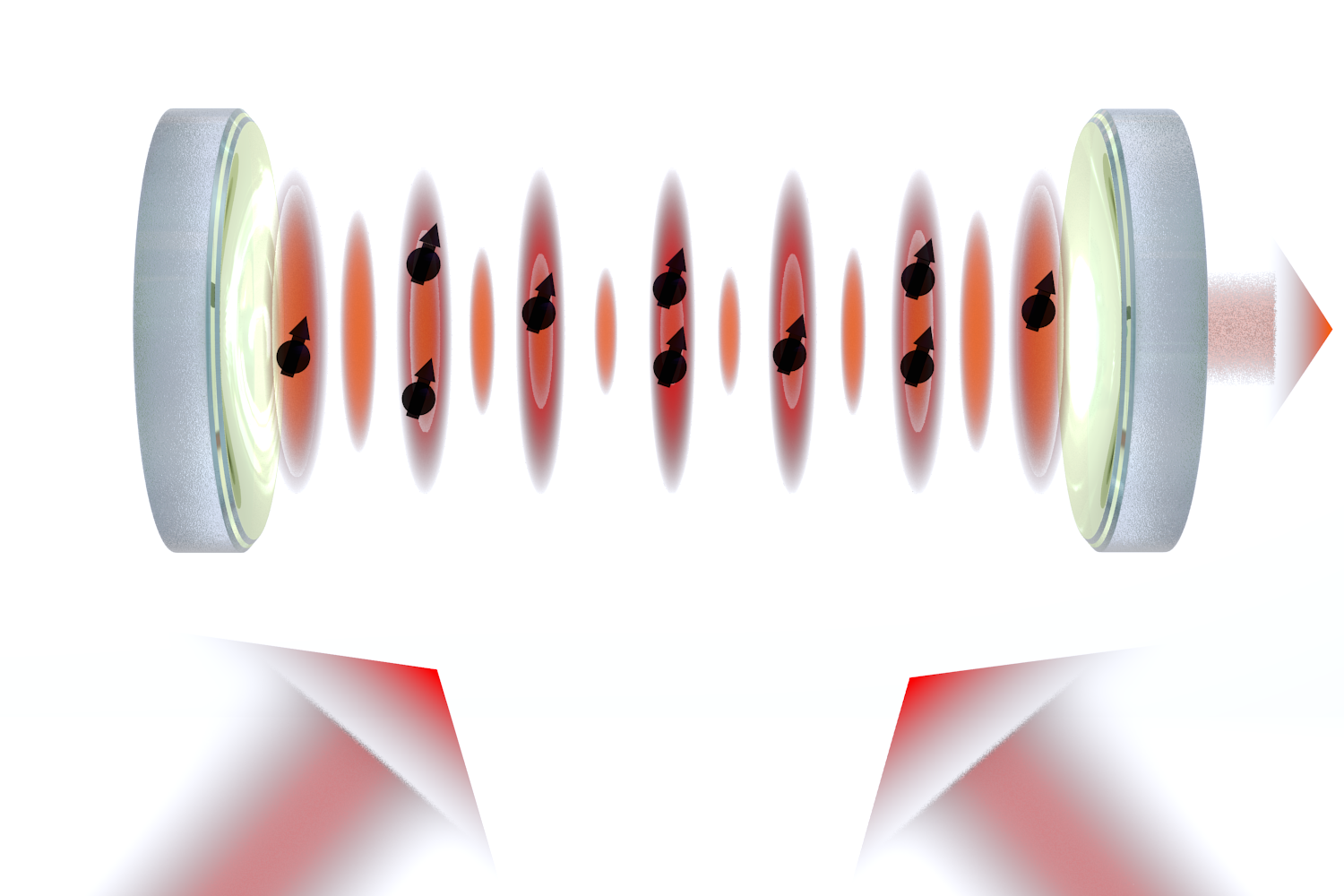}
    \caption{Scheme of an active optical atomic clock. Atoms (black) are trapped by an optical lattice at the magic wavelength (red) and coupled to the optical mode of an ultra-stable Fabry-Perot cavity (yellow).}
    \label{fig_LSR}
    \end{center}
\end{figure}

The idea of active optical clocks stems from superradiance, a quantum phenomenon initially described by Dicke in 1954~\cite{Dicke_1954}. When a large number $N$ of indistinguishable emitters are coupled to a single electromagnetic field, the constructive interference of possible decay paths between the fully excited state to the global ground state leads to an enhanced, collective emission called superradiance. Its intensity scales with $N^2$, while its duration is reduced by a factor $N$ with respect to uncoupled spontaneous emission~\cite{Gross1982}. 
Superradiance has been observed in various experimental media, such as quantum dots \cite{Scheibner_207_SR_quantum_dots, Tighineanu_2016_single_photon_SR_quant_dot}, light-harvesting complexes \cite{Doria_2018_photochem_control_SR}, %MD: the first article is a theory article (Celardo_2012_SR_photosynt)
or astrophysical phenomena \cite{Rajabi_2016_SR_astrophys, Blas_2020_quenching_photon_SR, Baryakhtar_2021_black_hole_SR}. %MD! this is theory:  and high-precision quantum sensing applications \cite{Koppenhofer_SR_quantum_sensing, BAZHENOV_2020_temp_quantum_sensor_SR}. 
%MD! It examines the concept of radiating gases as a single quantum-mechanical system and explores the resulting energy levels and their correlations. Dicke's focus is on the spontaneous emission of coherent radiation from these systems, especially in gases where the size is small compared to the radiation wavelength. 
In the case of emission in the radiofrequency, microwave, or far-infrared domain, coupling the emitters to a common field can be achieved by placing the emitters at a distance smaller than the emission wavelength $\lambda_0$~\cite{Skribanowitz_1973_superrad_opt_pumped_gas, superfluorescence1998}. 
%\textcolor{orange}{I would maybe switch the previous two sentences. Then, we would have a sentence about where in general superradiance has been observed. Then we would mention the radio-freq domain and how it can be achieved there, and then the next paragraph is about the optical domain with atoms... I don't know if it makes sense to you, but in my head it makes sense. :) But I let you decide.}
Regarding superradiant emission in the optical domain using atomic emitters, 
a dense sample with many atoms in a volume smaller than $\lambda_0^3$ would undergo decoherence preventing superradiant emission, 
so instead atoms are coupled to a single mode of an optical cavity~\cite{Haake_superrad_laser_1993}, as it is depicted in Fig. \ref{fig_LSR}.

\begin{figure}
    \begin{center}
    \includegraphics[width=9cm]{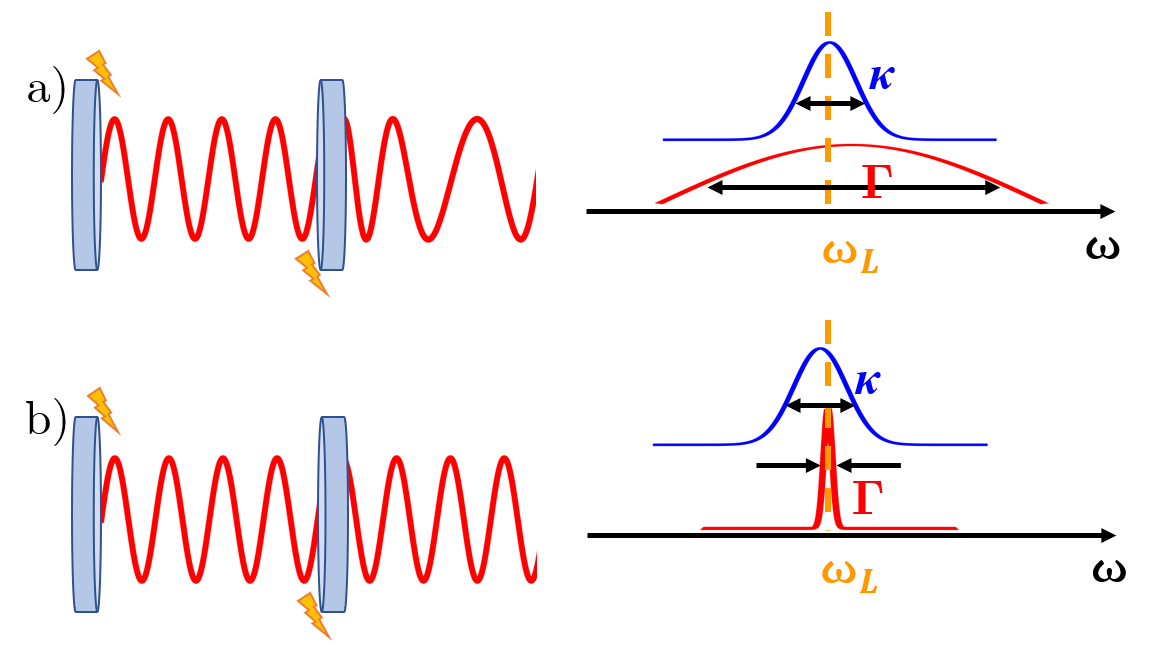}
    \caption{a) In the good-cavity regime, the laser output phase is dictated by the cavity, with fluctuations (symbolized by a lightning flash) on cavity mirrors notably affecting laser light properties.
    %In a good-cavity regime, the phase of the laser output is primarily determined by the cavity. Fluctuations (depicted as a lightning flash) affecting the cavity mirrors significantly influence the properties of the laser light. 
    b) In the bad-cavity regime, atoms retain the phase information, substantially reducing the impact of cavity perturbations.
    %Within the bad-cavity regime, the effects of cavity perturbations are markedly diminished, as the phase information is preserved by the atoms themselves. As a result, the frequency of the emitted light, and thereby the emitted light’s frequency, and thus the laser’s stability and accuracy, are dictated by the atoms rather than the cavity.
    }
    \label{fig_good_bad_cavity}
    \end{center}
\end{figure}

%MD! To couple atoms to a common electromagnetic field, the relative distance between individual emitters should be on the scale of the emission wavelength ${\lambda}_0$. 
%MD! Generally, although this condition is attainable for wavelengths in the microwave, radio-frequency, or even far-infrared transitions \cite{Skribanowitz_1973_superrad_opt_pumped_gas}, realizing it for dilute gases interacting with optical fields poses challenges. This proximity can cause energy level shifts that undermine the clock's accuracy.
In this case, the key figures are the cavity linewidth $\kappa$, the spontaneous emission rate $\gamma$ and the single-atom Rabi frequency $ 2g$. The build-up of quantum correlations that lead to a superradiance pulse is only possible when the collective vacuum Rabi frequency $\Omega = 2 g \sqrt{N}$ is much larger that $\kappa$ and $\gamma_\perp = \gamma/2 + 1/T_2$, where $T_2$ is the typical timescale of the other decoherence mechanisms in the system. 
Optical superradiance is usually realized in a regime where $\gamma\ll g$, so that photons are preferably emitted in the cavity mode, and where $g \ll \kappa$ so that the probability of emitted photons escaping the cavity exceeds the probability of their re-absorption by atoms. The system is then in the so-called ``bad-cavity regime'' illustrated in Fig.~\ref{fig_good_bad_cavity} and the coherence is stored in the atomic media rather than in the light field as is the case for usual, ``good-cavity'' lasers.

%it plays a crucial role for the use of superradiant systems as frequency references: in this case, 
%MD!Furthermore, to uphold the conditions conducive to the spontaneous emission regime and avoid stimulated emission, it is imperative to sustain $g \ll \kappa$. This inequality ensures that the probability of emitted photons escaping the cavity exceeds the probability of their re-absorption by atoms. In contrast to the operational dynamics of traditional lasers, where stimulated emission plays a central role, the system here is characterized as the "bad cavity regime". In this configuration, intentionally maintaining a low coupling strength ensures that emitted photons undergo only weak coupling with the cavity. 

%The introduction of atom repumping to sustain superradiance \cite{Haake_1993_superrad_laser},\textcolor{blue}{i'd like to remove the idea of repumping from here, and to remain more general, with only the idea of the superradiant *laser* // or to say that so far we discussed only superradiant pulses but now focus on a superradiant laser, a concept that has been introduced in -- or }
%The idea of renewing the atoms that have decayed to the ground state in order to maintain population inversion has led to the concept of superradiant lasers~\cite{Haake_superrad_laser_1993}. 
The concept of superradiant lasers stems from the process of replenishing atoms that have transitioned back to the ground state, maintaining population inversion~\cite{Haake_superrad_laser_1993}. 
In these systems, the renewing rate is of paramount importance, as a too-small rate cannot maintain a population inversion, while an excessively high rate induces too much decoherence and can destroy the collective dipole of the system.
Their very high potential as optical frequency references \cite{Chen2009, Meiser_prospects_for_mHz_lw_laser_2009} has paved the way for the development of active optical atomic clocks, that can be seen as an optical counterpart to the masers. 
Especially, it has been predicted that the $N^2$ scaling of the output power could lead to values in the picowatt range on a millihertz linewidth transition. 
%Especially, it has been predicted that the continuous output power would scale as $N^2$, with a total output power in the picowatt range on a millihertz linewidth transition, 
The superradiant emission linewidth could be as low as $C \gamma$, where $C$ is the cooperativity factor 
\begin{equation}
    C = \frac{(2 g)^2}{\kappa\gamma},
\end{equation}    
laying the foundations for oscillators with linewidths of a few mHz and fractional frequency instabilities in the $10^{-18}$ range at one second. 
Operation in the bad-cavity regime ensures that superradiant photons are minimally impacted by the intrinsic frequency instability of the cavity. The change of the superradiant laser output frequency due to a cavity frequency change is  characterized by the cavity pulling coefficient 
\begin{equation}
    P = \mathrm{d}f/\mathrm{d}f_\mathrm{cav} =  2 \gamma_\perp /(2\gamma_\perp+\kappa)\approx 2 \gamma_\perp/\kappa
\end{equation}
that can be in the $10^{-5}$ range.
%\textcolor{blue}{it is probably worth introducing the idea of cavity pulling, its simplified expression and accessible values -- maybe this paragraph could go a bit after, since superradiant *lasers* have not been introduced yet, and the low cavity pulling could go with performances perdictions}

%\textcolor{red}{write some results and perspectives from Meiser's work!}
%MDThe idea of active optical clocks, that uses direclty the light emitted by an atomic ensemble as an ultra-stable signal, emerged from two seminal papers: the description of a superradiant laser~\cite{Haake_1993_superrad_laser}, and its application for a narrow-linewidth laser~\cite{Meiser_prospects_for_mHz_lw_laser_2009}, that can play the role of a frequency reference. 

%While masers, often calibrated on Cs clocks, serve as highly reliable and stable tools for radio-frequency timescales over both short and long term, an equivalent optical counterpart is presently lacking. The best short-term local oscillators today, optical Fabry-Perot cavities, exhibit unpredictable long-term drifts. To surpass the existing state-of-the-art and address issues related to local oscillator frequency instability, an intriguing concept is to develop an active optical atomic clock.

Two main strategies are currently investigated to create a superradiant active optical clock: optical lattice superradiant lasers~\cite{Meiser_prospects_for_mHz_lw_laser_2009}, and atomic beam superradiant lasers~\cite{Chen2009}. %\textcolor{blue}{they can be distinguished by the way they renew their excited atoms? -- find a nice way to say this}
%\textcolor{purple}{These methods are primarily differentiated by their distinct mechanisms for replenishing their excited atoms.}
These methods are primarily differentiated by their distinct mechanisms to maintain inversion population.

\subsection{Optical lattice superradiant lasers}
%\textcolor{blue}{At the moment, optical lattice SR lasers do not exist yet as is, so this is just an *idea* yet. Emphasize the idea. In optical lattice lasers, atoms are confined within an optical lattice operating at a magic wavelength~\cite{Maier_14_SR_laser_magic_wavelength} to operate in the Lamb-Dicke regime. Trapped atoms are continuously pumped in order to maintain inversion population for continuous lasing. }
%\textcolor{purple}{In the realm of superradiant active optical clocks, the concept of optical lattice lasers stands out as a promising yet theoretical idea. Within this framework, atoms would be confined in an optical lattice operating at a 'magic wavelength', enabling operation in the Lamb-Dicke regime. This setup envisions trapped atoms being continuously pumped to maintain population inversion, essential for sustained lasing.}
%Superradiant lasers based on atoms trapped in optical lattices stand out as a promising yet theoretical idea. 
Superradiant lasers based on atoms trapped in optical lattices stand out as a promising concept for ultra-stable oscillators. 

Within this framework, atoms would be confined in an optical lattice operating at a so-called magic wavelength and operate in the Lamb-Dicke regime~\cite{Katori_2003_magic_lambda}. In this configuration, trapped atoms are continuously pumped to maintain population inversion, essential for sustained lasing. 
After the seminal paper for optical lattice superradiant lasers in 2009~\cite{Meiser_prospects_for_mHz_lw_laser_2009}, the first proof of principle for optical lattice superradiant lasers was obtained at JILA~\cite{Bohnet2012_steady_state_superrad_less1_intracav_photon} using Raman dressed states of ${}^{87}\mathrm{Rb}$ repumped to reach a steady-state. The $N^2$ scaling of the output power was demonstrated, and the conditions for the stable lasing operation in their effective 3-level system were explored in~\cite{Bohnet_2012_relax_osc_stab_cavity_feedback_superrad_raman_laser}. A scheme for a four-level active optical clock based on ${}^{87}\mathrm{Rb}$ has also been proposed~\cite{Zhang_2013_4level_lasing_system}. 

Superradiant emission directly on a relatively narrow optical transition was demonstrated on the 8~kHz-wide ${}^1\mathrm{S}_0 \leftrightarrow {}^3\mathrm{P}_1$ transition of ${}^{88}\mathrm{Sr}$ at JILA~\cite{Norcia_cold_Sr_laser_superrad_crossover_regime} and at UCPH~\cite{Kristensen2023}, and on the 375~Hz-wide ${}^1\mathrm{S}_0 \leftrightarrow {}^3\mathrm{P}_1$ transition of ${}^{40}\mathrm{Ca}$ in Hamburg~\cite{Laske2019}. While atoms are trapped in a 1D optical lattice at JILA and Hamburg, at UCPH they are prepared in a Magneto-Optical Trap (MOT) inside a Fabry-Perot cavity and then released in free-fall to produce superradiance in the cavity. 
%\textcolor{blue}{we probably should discuss a bit more the three setups: at the moment it is not obvious that UCPH experiment is "untrapped", at JILA they are trapped in a magic lambda lattice, in Hamburg I forgot.} \textcolor{orange}{See below in purple my suggestion how to mention this. I think it's ok if we just mention it in the end, because we first focus on their achievments. Maybe we should say more about what has been achieved in Hamburg?}
The JILA and the UCPH groups managed to reach a steady state regime for a few milliseconds by repumping the atoms. At JILA, the emission ceases after the emission of about 35 photons per atom because of repump-induced heating, while at UCPH it is limited by the free-fall of the atoms that leave the cavity volume after the emission of approximately one photon per atom. 
Linewidths measurements were performed by UCPH group showing a subnatural linewidth of 820~Hz, Fourier-limited by the emission duration~\cite{Kristensen2023}. %\textcolor{purple}{It should be noted that, while JILA and Hamburg have achieved superradiant lasing with atoms trapped in an optical lattice at the magic wavelength, UCPH has successfully produced stable superradiant pulses utilizing a free-falling cloud of cold atoms coupled to a low finesse cavity.}

The JILA group also demonstrated pulsed superradiant emission on the ${}^1\mathrm{S}_0 \leftrightarrow {}^3\mathrm{P}_0$ clock transition of ${}^{87}\mathrm{Sr}$~\cite{Norcia_superrad_on_mHz_linwidth_Sr_clock_trans}. A heterodyne beatnote between superradiant pulses and a stable laser was used to extract the average frequency of each pulse and perform frequency stability measurements of the succession of pulses. A fractional frequency of $6\times 10^{-16}\tau^{-1/2}$ (for $\tau$ larger than the cycle time of 1.1~s) has been reported~\cite{Norcia_freq_measur_of_superrad_from_Sr_clock_trans}, assessing a promising long-term frequency stability. %However, because of the intermittent emission, there is no phase continuity between the pulses and short-term stability over the pulse duration has not been evaluated. 
%\textcolor{purple}{Due to the intermittent interrogation the system faces two challenges: firstly, there is no phase continuity between the pulses; secondly, the system is unable to provide short-term stability. These issues both impact the overall performance of the system.}
The system however faces two challenges due to the intermittent emission: firstly, there is no phase continuity between the pulses; secondly, the short-term stability over the pulse duration has not been evaluated. %\textcolor{blue}{These issues both impact the overall performance of the system.}

%\textcolor{blue}{cautious about repumping scheme + cite JILA}
One of the main current challenges for superradiant optical clocks is therefore the realization of a true continuous operation. 
A prevalent solution is to use an incoherent pump~\cite{Haake_superrad_laser_1993, Meiser_prospects_for_mHz_lw_laser_2009, Henschel_cavity_QED_ultracold_ensemble_on_chip_2010}, efficiently repumping atoms from the ground to the excited state during superradiant emission. This approach ensures sustained population inversion without introducing coherence between the pump and lasing mode, therefore avoiding amplified phase diffusion and intensity fluctuations. Relaxation oscillations and stable behavior of the effective 3-level superradiant laser based on dressed states of ${}^{87}\mathrm{Rb}$ has been investigated in~\cite{Bohnet_2012_relax_osc_stab_cavity_feedback_superrad_raman_laser}. It has been shown that the repumping rate could be tuned to control relaxation oscillations. 
UCPH also investigated the conditions for stable lasing in their system. They observed strong oscillations of the output power, consistent with an excessive gain, when using an effective 3-level system, and could reach a promising stable emission regime with an effective 4-level scheme. 
% and demonstrated that repumping on an effective 3-level scheme led to instabilities and oscillatory behavior of the output power, \textcolor{blue}{characteristic of a system with a too large gain, while the gain is hard to tune on a 3-level system. On the contrary, their effective 4-level scheme led a a stable and promising behavior.}
However, the lifetime of atoms inside the cavity is anyway limited, and intense experimental efforts are currently dedicated to the issue of guiding new cold atoms to the cavity in order to compensate for atom lost during superradiant emission.

%\textcolor{blue}{this transition is bad, find a way to improve it} 
%Yet, several proposals directly tackle this difficulty by considering atomic beam superradiant lasers.
%\textcolor{purple}{In light of these challenges, several innovative proposals have emerged, focusing specifically on atomic beam superradiant lasers. These proposals directly confront the difficulties associated with repumping and collective dipole disruptions, representing a pivotal step forward in the development of more effective and reliable superradiant lasing systems.}

In light of these challenges, several proposals tackle directly the two issues of collective dipole continuation and atomic lifetime  by investigating the feasibility of atomic beam superradiant lasers and their expected properties.

\subsection{Atomic beam superradiant lasers} 
In atomic beam superradiant lasers, a continuous stream of atoms, pre-excited to the upper lasing state, traverses the cavity where they produce superradiance~\cite{Chen2009}. In these systems, atoms are untrapped and experience minimal light shifts, but other challenges arise. 
In particular, the finite transit time $\tau_\mathrm{tr}$ of atoms through the cavity affects the boundaries of the continuous superradiant lasing regime~\cite{Yu_Chen_2008_atom_beam_laser_theory, LPL_beam_cw_SRL_2023}.
The relevant regime for metrological applications is obtained when $\tau_\mathrm{tr} \gg \kappa/(N g^2) $, the typical timescale necessary for the formation of the collective dipole.  In this case, the output linewidth is given by $C \gamma$.
%$C \gamma$ \textcolor{blue}{when N is large, the time needed for the synchronisation of the atoms is small, so the // we want a long transit time, so that atoms can reach the steady-state inside the cavity // transit time long enough for them to sync}
%the output properties of the superradiant laser and leads to a broadening of its linewidth when $\tau_\mathrm{tr} \ll \gamma$~\cite{Yu_Chen_2008_atom_beam_laser_theory,LPL_beam_cw_SRL_2023}. 
%The inhomogeneous broadening related to a first-order Doppler shift resulting from the atomic motion along the cavity axis also affects the superradiant transition threshold. 
The inhomogeneous broadening related to a first-order Doppler shift resulting from the atomic motion along the cavity axis leads to additional linewidth broadening.

%The impact of finite atom transit time on laser linewidth has been investigated in~\cite{Yu_Chen_2008_atom_beam_laser_theory}, \textcolor{blue}{examining the spectrum of fluctuations in the output field, and exploring how pumping statistics affect the output field <-- this sentence is exactly extracted from their abstract ! it should be rephrased!}. There, the authors assumed a homogeneously broadened atomic medium, thereby neglecting the first-order Doppler shift resulting from the atomic motion along the cavity axis. \textcolor{blue}{conclusion from \cite{Yu_Chen_2008_atom_beam_laser_theory}?} 
A realistic scheme has been proposed in~\cite{Liu_2020_rugged_laser}, showing that linewidths in the mHz regime were possible with a simple and robust scheme using thermal atoms. One of the main challenges of these systems is the relatively high flux of atoms that is necessary to initiate and sustain superradiance, but remarkable continuous guidance of ${}^{88}\mathrm{Sr}$ with a high phase-space density and sub-${\upmu}\mathrm{K}$ radial temperature has been demonstrated recently~\cite{Schreck_continuous_guided_beam, Chen_2022} and could lead to precursors of cold superradiant beam lasers.

%The initial theory proposing the generation of an atom beam laser was outlined in \cite{Yu_Chen_2008_atom_beam_laser_theory}, discussing the impact of finite atom transit time on laser linewidth, examining the spectrum of fluctuations in the output field, and exploring how pumping statistics affect the output field. There, the authors assumed a homogeneously broadened atomic medium, thereby neglecting the first-order Doppler shift resulting from the atomic motion along the cavity axis.

\subsection{Hybrid superradiant laser scheme}

An alternative approach suggests the combination of the optical lattice laser and the atomic beam laser, dynamically transferring trapped atomic ensembles into and out of the cavity~\cite{kazakov2015active_opt_freq_st}. As previously mentioned, the significance of adequate transit time through the cavity persists. 
A more specific proposal involves introducing atoms already in the upper lasing state into the cavity while atoms from the previous ensemble are still emitting ~\cite{Kazakov_2013_sequential_coupling}.
%using a moving red-detuned one-dimensional optical lattice operating at the magic wavelength \cite{Kazakov_2013_sequential_coupling}. There, the atoms are confined to a blue-detuned stationary optical lattice at the magic wavelength within the cavity. The atomic ensembles are prepared in the upper lasing state outside the cavity, avoiding perturbations caused by ac Stark shifts. To suppress the second-order Doppler shift, a constant velocity is maintained for the moving lattice as atoms traverse the cavity waist. 
Sequentially introduced into the cavity, the inverted atomic ensembles stimulate the atoms of each subsequent ensemble to emit photons into the cavity mode primarily through stimulated emission, thereby sustaining optical phase coherence. 
%The authors also explore the duration of the lasing cycle and its effects on intracavity field phase fluctuations. Additionally, the study reveals the robustness of the phase relay process against fluctuations in the introduced atom number within the cavity, positioning it as a promising approach for developing an active optical frequency standard. 
%Various proposals for the hybrid laser scheme integration are discussed in the theoretical study by \cite{kazakov2015active_opt_freq_st}. One suggestion is to employ a moving optical lattice tightly focused on the lasing region. %Another intriguing proposal involves the preparation of atoms in the upper lasing state outside the cavity, followed by their introduction into a blue-detuned optical lattice operating at the magic wavelength. -> This one is actually what is described at the beginning of this paragraph!

Extensive theoretical and experimental investigations have concentrated on pulsed and quasi-continuous superradiant emission. Nevertheless, achieving continuous lasing remains a fundamental objective with broad applications. Merging the advantages of the superradiant optical lattice laser and the superradiant atomic beam laser, the combined hybrid lasing scheme offers the potential to overcome the limitations associated with pulsed superradiant lasing, potentially paving the way for truly continuous operation.

\subsection{Limitations of active optical atomic clocks}
%\textcolor{orange}{I think the title should be changed to be more general, for example, Limitations or possible limitations of active opt. atomic clocks}

%\textcolor{blue}{I took this from before, find where to put it:}
%\textcolor{blue}{In addition to the constrained lifetime of atoms within the cavity, resulting in pulsed emission, and losses due to atom heating caused by continuous repumping, another challenge affecting the performance of the optical lattice superradiant laser is the interference from first-order Zeeman and vector light shifts \cite{kazakov2015active_opt_freq_st}. While the latter issue can be alleviated through magnetic field stabilization, the loss of atoms caused by continuous repumping can be tackled by shelving atoms into a long-lived state. }

%\textcolor{blue}{H. Ritsch and co-workers have studied DDI in a magic lattice\cite{Maier_14_SR_laser_magic_wavelength}}

The performance of superradiant lasers may be constrained by the difficulty of maintaining uniform coupling of atoms to a common field and efficiently pumping the atoms without perturbing the system. Additionally, dipole-dipole interactions between the atoms can lead to decoherence in the collective dipole and generate phase noise, ultimately limiting the laser's frequency stability. %These limitations have been theoretically explored in two different papers, each addressing one of these specific challenges, and their results will be discussed below.

%H. Ritsch and his co-workers' study on superradiant lasers focuses on dipole-dipole interactions between atoms within the cavity, identifying them as a major noise source in low-density lattices \cite{Maier_14_SR_laser_magic_wavelength}. 
A particular study concentrates on dipole-dipole interactions between atoms within the cavity, identifying them as a major noise source in low-density lattices \cite{Maier_14_SR_laser_magic_wavelength}. 
%Utilizing an idealized pumping method to reduce decoherence, while disregarding light shifts from pump lasers, they found that optimal laser linewidth is achieved at moderate pump strengths, with increased perturbations from collective interactions at higher pump strengths and closer atomic proximity. Their results revealed that laser frequency changes due to cavity interactions remain within an atomic linewidth despite fluctuations. They also found that in highly dense atomic assemblies, superradiant decay notably broadens the laser linewidth and increases its sensitivity to cavity fluctuations. 
Utilizing an idealized pumping method to reduce decoherence, while disregarding light shifts from pump lasers, they found that optimal laser linewidth is achieved at moderate pump strengths and is robust against cavity fluctuations even when these significantly exceed the laser linewidth. However, superradiant decay of a highly dense atomic assembly notably broadens the laser linewidth and increases its sensitivity to cavity fluctuations.
%Utilizing an idealized pumping method to reduce decoherence, while disregarding light shifts from pump lasers, they found that optimal laser linewidth is achieved at moderate pump strengths and is robust to cavity fluctuations even of the order of the cavity linewidth. However, superradiant decay of a highly dense atomic assembly notably broadens the laser linewidth and increases its sensitivity to cavity fluctuations.

%Another recent study that specifically disregarded dipole-dipole interactions between distinct atoms and also excluded consideration of the collective heat bath, investigated the variation of laser linewidth on different laser parameters
%with parameters such as repumping rate, cooperativity factor, and atomic dephasing 
%in both homogeneous and inhomogeneous systems \cite{Kazakov_2022}.  
Another recent study investigated the variation of laser linewidth on different laser parameters in both homogeneous and inhomogeneous systems \cite{Kazakov_2022}.  
This study demonstrated that the linewidth of a superradiant laser is robust to the incoherent repumping rate, but for systems with inhomogeneous atomic dephasing, both the minimum achievable linewidth of the laser and the number of intracavity photons are affected by the atoms dephasing rate. 

 The same study focused on an inhomogeneous system of an atomic ensemble interacting with a single cavity mode to examine the attainable frequency stability of an active optical atomic clock. It has been demonstrated that the ultimate stability of an incoherently pumped active optical atomic clock can achieve a level of stability equivalent to the quantum projection noise (QPN)-limited stability of a passive optical atomic clock at the same atom number for both types of clocks.
%In a more comprehensive analysis of the fractional frequency stability, the study considered the dephasing of the atomic transition due to Raman scattering of photons from the optical lattice potential, as well as the imperfections of the local oscillator in the context of passive clocks. 
Instabilities of ${\sigma}_{y}(\tau) \approx 3.7 \times 10^{-18} / \sqrt{\tau}$ should be reachable, with the short-term stability constrained by the shot noise of the active clock. This demonstrates that the short-term stability of an active optical atomic clock can either match or even significantly surpass the stability of an ideal, zero dead-time, QPN-limited passive optical atomic clock with the same number of atoms as the considered active clock. 
%However,  there is a possibility of suppressing the Dick effect in specific scenarios. This suppression has been observed when comparing two different atomic species within an optical lattice, as supported by \cite{Schioppo_2017_2_ensembles_clock_reduced_dick_eff}, or when comparing two different clocks at the same transition \cite{Takano_2016_2_clocks_reduced_Dick_eff}. 
On the other hand, the long-term stability of active optical atomic clocks will be constrained by the phase diffusion of the superradiant signal at the cavity output.

In addition to the constrained lifetime of atoms within the cavity, resulting in pulsed emission, and losses due to atom heating caused by continuous repumping, another challenge affecting the performance of the optical lattice superradiant laser is the interference from first-order Zeeman and vector light shifts \cite{kazakov2015active_opt_freq_st}. In contrast to passive clocks, where the Zeeman effect induced by a bias magnetic field can be suppressed by averaging Zeeman transitions with opposite shifts, alternating from one interrogation cycle to the next, active clocks may operate on two Zeeman transitions simultaneously. However, this potential issue can be effectively mitigated by configuring the magnetic field in a manner that keeps the splitting between these two transitions significantly smaller than the cavity linewidth, allowing the system to remain within the bad-cavity regime. Furthermore, the splitting should be considerably larger than the pumping rate, ensuring that there is no reduction in the achievable stability, as discussed in \cite{Kazakov_2022}. On the other hand, the loss of atoms caused by continuous repumping can be tackled by shelving atoms into a long-lived state.
%In contrast to passive clocks, where the Zeeman effect induced by a bias magnetic field can be suppressed by averaging Zeeman transitions with opposite shifts, alternating from one interrogation cycle to the next, active clocks may operate on two Zeeman transitions simultaneously. However, this potential issue can be effectively mitigated by configuring the magnetic field in a manner that keeps the splitting between these two transitions significantly smaller than the cavity linewidth, allowing the system to remain within the bad-cavity regime. Furthermore, the splitting should be considerably larger than the pumping rate, ensuring that there is no reduction in the achievable stability, as discussed in \cite{Kazakov_2022}.
%Ultimately,  the superradiant clock is not practically usable on its own, due to its low output power. Therefore, an additional laser is required to be phase-stabilized onto the superradiant signal. In this context, the short-term stability is constrained by the shot noise of the active clock, which defines the bandwidth of the phase-locking process. On the other hand, the long-term stability of active optical atomic clocks will be constrained by the phase diffusion of the superradiant signal at the cavity output, as noted in \cite{Kazakov_2022}.

Finally, unlike passive clocks, active clocks can sustain their accuracy even when the frequency of the cavity resonance marginally strays from the atomic transition. This characteristic makes active clocks potential candidates for applications outside controlled laboratory environments, where it is challenging to suppress environmental noise. This perspective would require a thorough examination of systematic effects.
%Furthermore, the stability of active optical atomic clocks can surpass that of passive clocks, suggesting a promising future for the advancement of active optical atomic clocks.

\section{Conclusion}
In this paper we presented a comprehensive examination of the development and potential of superradiant active optical atomic clocks. These timekeepers represent a significant leap forward from the current passive optical atomic clocks. The core advantage of active clocks lies in their inherent robustness against cavity instabilities, as their emitted frequency is defined by atomic transitions rather than being reliant on external factors.

We have explored various strategies and advancements in the field, such as optical lattice superradiant lasers and atom beam lasers. These approaches have shown promising results in achieving continuous operation and overcoming the limitations of passive clocks. Notably, the ability of active clocks to remain stable even when the frequency of the cavity resonance deviates slightly from the atomic transition marks them as more adaptable to environments outside controlled laboratory settings. Furthermore, their superior stability makes them well-suited for advanced applications in space-based navigation and gravitational wave detection.

Despite the progress, several challenges remain. The difficulty of achieving continuous lasing or managing atom heating due to continuous repumping are among the primary concerns. However, the ongoing theoretical and experimental efforts in this domain indicate a promising future for these clocks. The potential of active optical atomic clocks to achieve fractional frequency instabilities in the $10^{−18}$ range at one second is particularly noteworthy.

In conclusion, the advancement of active optical atomic clocks holds immense promise for the fields of metrology, geodesy, and fundamental physics. Their development not only represents a technological breakthrough but also opens the door to new scientific discoveries and applications, ranging from probing fundamental constants to detecting subtle cosmic phenomena.

\section*{Acknowledgements}
The authors would like to thank Clément Lacroûte, Emmanuel Klinger and Vincent Giordano for their fruitful comments and careful reading of the manuscript.

\section*{References}
\bibliography{references}
%\begin{thebibliography}{9}
%\bibitem{biblio.bib} IOP Publishing is to grateful Mark A Caprio, Center for Theoretical Physics, Yale University, for permission to include the {\tt iopart-num} \BibTeX package (version 2.0, December 21, 2006) with  this documentation. Updates and new releases of {\tt iopart-num} can be found on \verb"www.ctan.org" (CTAN). 
%\end{thebibliography}

\end{document}